\documentclass[aps,prd,preprint,showpacs,nofootinbib]{revtex4}
\usepackage{epsfig}
\usepackage{amsmath}
\usepackage{amssymb}
\usepackage{amsbsy}
\usepackage{amsfonts}
\usepackage{graphics}
\usepackage{latexsym}
\usepackage{mathrsfs}
\usepackage{dcolumn}
\usepackage{wasysym}
\usepackage[dvips]{color}
\usepackage{graphicx}
\usepackage[figuresright]{rotating}

\begin{document}

\begin{raggedright}
DESY 08-112\\
Edinburgh 2008/11
\end{raggedright}

\title{The electric dipole moment of the nucleon from
  simulations at imaginary vacuum angle theta}

\author{R.~Horsley$^{1}$,
        T.~Izubuchi$^{2,3}$, 
        Y.~Nakamura$^{4}$, D.~Pleiter$^{4}$, P.E.L.~Rakow$^{5}$,
        G.~Schierholz$^{4}$ and J.~Zanotti$^{1}$} 

\affiliation{\vspace*{0.3cm} 
$^1$ School of Physics, University of Edinburgh,
  Edinburgh EH9 3JZ, UK\\
$^2$ Institute for Theoretical Physics, Kanazawa University,
  Kanazawa, 920-1192 Japan\\ 
$^3$ RIKEN-BNL Research Center, Brookhaven National Laboratory, Upton, NY
  11973, USA\\
$^4$  Deutsches Elektronen-Synchrotron DESY, John von Neumann-Institut
  f\"ur Computing NIC, 15738 Zeuthen, Germany\\
$^5$ Theoretical Physics Division, Department of Mathematical
  Sciences, University of Liverpool, Liverpool L69 3BX, UK}

%\author{S.~Aoki$^{1,2}$, R.~Horsley$^{3}$,
%        T.~Izubuchi$^{1,4}$, 
%        Y.~Nakamura$^{5}$, D.~Pleiter$^{5}$, P.E.L.~Rakow$^{6}$,
%        G.~Schierholz$^{5}$ and J.~Zanotti$^{3}$} 
%
%\affiliation{\vspace*{0.3cm} 
%$^1$ RIKEN-BNL Research Center, Brookhaven National Laboratory, Upton, NY
%  11973, USA\\
%$^2$ Graduate School of Pure and Applied Sciences, University of Tsukuba,
%Tsukuba, 305-8571 Japan\\
%$^3$ School of Physics, University of Edinburgh,
%  Edinburgh EH9 3JZ, UK\\
%$^4$ Institute for Theoretical Physics, Kanazawa University,
%  Kanazawa, 920-1192 Japan\\ 
%$^5$  Deutsches Elektronen-Synchrotron DESY, John von Neumann-Institut
%  f\"ur Computing NIC, 15738 Zeuthen, Germany\\
%$^6$ Theoretical Physics Division, Department of Mathematical
%  Sciences, University of Liverpool, Liverpool L69 3BX, UK}

%\author{QCDSF-DIK Collaboration}

\vspace*{1cm}

\begin{abstract}
We compute the electric dipole moment of proton and neutron from lattice QCD
simulations with $N_f=2$ flavors of dynamical quarks at imaginary vacuum angle
$\theta$. The calculation proceeds via the $CP$ odd form factor $F_3$. A novel
feature of our calculation is that we use partially twisted boundary
conditions to extract $F_3$ at zero momentum transfer. As a byproduct, we test
the QCD vacuum at nonvanishing $\theta$.     
%We compute the electric dipole moment of proton and neutron in lattice QCD 
%for $N_f=2$ flavors of dynamical quarks at nonvanishing imaginary vacuum angle
%$\theta$. For the neutron we find $\displaystyle d_N^n = -4.9(5) \, \times
%10^{-15} \, \theta \,e \,\mbox{cm}$. Combined with the experimental limit on
%the electric dipole moment this leads to $|\theta|<6 \times 10^{-12}$.
\end{abstract}

\pacs{11.30.Er,11.15.Ha,12.38.Gc,13.40.Gp}

\maketitle

%\newpage
\section{Introduction}
Current measurements of $CP$ violating processes in the $K$ and $B$ meson
sector would suggest that the phase of the CKM matrix provides a complete
description. However, the baryon asymmetry of the universe cannot be described
by this phase alone, suggesting that there are additional sources of $CP$
violation awaiting discovery. 

QCD allows for a gauge invariant extra term in the action that is odd under
$CP$ transformations, 
\begin{equation}
S \rightarrow S + i\,\theta\,Q ,
\end{equation}
where $Q$ is the topological charge. Hence, there is the possibility of strong
$CP$ violation arising from a nonvanishing vacuum angle $\theta$. The presence
of $CP$ violating forces implies a permanent electric dipole moment of the
proton and neutron. This attribute is also deeply related to the question of
baryon asymmetry of the universe~\cite{Trodden}.  

The current experimental bound on the electric dipole moment of the neutron
is~\cite{Baker} 
\begin{equation}
|d_N^n| < 2.9 \,\times \, 10^{-13} \, e\,\mbox{fm} 
\label{ub}
\end{equation}
($\displaystyle 1\, \mbox{fm} = 10^{-13} \, \mbox{cm}$). Combining this bound
with theoretical estimates of $d_N/\theta$ allows us to derive an upper bound
on the value of $|\theta|$. Current estimates from QCD sum
rules~\cite{Pospelov} and chiral perturbation theory~\cite{Borasoy} give
$\displaystyle |\theta| \lesssim (1 - 3) \times 10^{-10}$. This anomaly is
known as the strong $CP$ problem.  

With the increasingly precise experimental efforts to observe the
electric dipole moment of the neutron~\cite{Harris}, it is important to have a
rigorous calculation directly from QCD. 

In this paper we present a calculation of $d_N$ in units of $\theta$ with
$N_f=2$ flavors of dynamical quarks using the lattice regularization. The
novel feature of our work is that the simulations are performed directly at
nonvanishing vacuum angle $\theta$, in contrast to previous lattice
studies~\cite{Shintani1,Shintani2,Berruto,Shintani3} (with the exception
of~\cite{Izubuchi}), which rely on reweighting correlation functions with
topological charge that would otherwise vanish. The calculation becomes
feasible if $\theta$ is rotated to purely imaginary
values~\cite{Bhanot,Azcoiti,Imachi}. We expect from our 
method a much enhanced signal to noise ratio. In addition, we may hope to gain
some insight into the dynamics of the $\theta$ vacuum. For a recent review on
this matter see~\cite{Vicari}.   

\section{The action}

The vacuum angle $\theta$ can be rotated into the mass term, in the continuum
and on the lattice~\cite{Seiler,Kerler}. This results in the fermionic action
\begin{equation}
S_F=\bar{\psi}\left\{D + [\cos(\theta/N_f) +
  i\, \sin(\theta/N_f)\, \gamma_5]\,m\right\}\psi,  
\label{fa}
\end{equation}
where summation over space-time coordinates $\{\vec{x},t\}$ and quark flavors
is understood, and $D$ is the massless Dirac operator. For simplicity,
we shall write in the following 
\begin{equation}
S_F=\bar{\psi}\left\{D + \bar{m} +
  i\, (\bar{\theta}/N_f)\, \gamma_5\, \bar{m} \right\}\psi,  
\end{equation}
with 
\begin{equation}
\begin{split}
\bar{m} &= \cos(\theta/N_f)\, m ,\\
\bar{\theta} &= \tan(\theta/N_f)\, N_f .
\end{split}
\label{bar}
\end{equation}

We use clover fermions with $N_f=2$ flavors of degenerate quarks. Taking into
account that chiral symmetry is violated, we then have 
\begin{equation}
S_F=\bar{\psi}\left\{D + \bar{m} + i\, (\theta_R/2)\, Z_m^S Z_P \,
  \gamma_5 \, \bar{m}\right\}\psi,  
\label{action}
\end{equation}
where $Z_m^S$ is the {\em singlet} renormalization constant of
the vector Ward identity ($VWI$) quark mass and $Z_P$ that of the pseudoscalar
density, and $\theta_R$ is the renormalized vacuum angle,
\begin{equation}
\theta_R = \left(Z_m^S Z_P\right)^{-1} \, \bar{\theta} .
\end{equation}
It can be shown that $Z_P$ is the same in both the singlet and nonsinglet case,
while $Z_m$ is not~\cite{qcdsf}. Note that $Z_m^S Z_P$ is scale independent,
and in the continuum limit $Z_m^S Z_P = 1$ and $\theta_R = \bar{\theta}$.

\section{The simulation}

The simulations are performed with the Iwasaki gauge action and said clover
fermions with $c_{SW}=1.47$ on $16^3 \, 32$ lattices at $\beta=2.1$,
$\kappa=0.1357$~\cite{cppacs}. The (massless) Dirac operator $D$ in
(\ref{action}) is evaluated at $\kappa_c=0.138984$. The bare mass is taken to
be $a\bar{m} = 1/(2\kappa) - 1/(2\kappa_c)$, independent of $\theta$. The
resulting pion mass is $m_\pi/m_\rho \approx 0.8$, corresponding to a quark
mass of $m\approx m_s$, $m_s$ being the strange quark mass. In~\cite{cppacs} 
the lattice spacing in the chiral limit was estimated at $a \approx 0.11$ fm,
using the $\rho$ mass to set the scale.

The vacuum angle $\bar{\theta}$ is taken to be purely imaginary, 
\begin{equation}
\bar{\theta} = -i\, \bar{\theta}^{I} , \quad \bar{\theta}^{I} \in
\mathbb{R} ,
\label{choice}
\end{equation} 
resulting in the action
\begin{equation}
S_F=\bar{\psi}\left\{D + \bar{m} + (\bar{\theta}^{I}/2) \,
  \gamma_5 \, \bar{m}\right\}\psi.
\end{equation}
The simulations are done at $\bar{\theta}^{I} = 0, 0.2$, $0.4$, $1.0$ and
$1.5$, where we have collected $9000$, $9000$, $7000$, $6000$ and $6000$
trajectories of length one, respectively.

We use the highly optimized HMC algorithm of the QCDSF
Collaboration~\cite{AliKhan} for updating the gauge field. After integrating
out the Grassmann fields, the action reads
\begin{equation}
S[U,\phi^\dagger,\phi] = S_{\rm G}[U] + S_{\rm det}[U] + \phi^\dagger
(Q^\dagger Q)^{-1}\phi, 
\label{standard}
\end{equation}
where $S_G[U]$ is the Iwasaki gauge action, $\phi^\dagger$ and $\phi$ are
pseudo\-fermion fields, and  
\begin{equation}
\begin{split}
S_{\rm det}[U] &= -2\, {\rm Tr}\, \log\left(1+T_{\rm
  oo}+\frac{\hat{\theta}^{I}}{2} \, \gamma_5 \right),\\
Q &= \left(1+T + \frac{\hat{\theta}^{I}}{2}\, \gamma_5\right)_{\rm ee} -
  M_{\rm eo} 
  \left(1+T +\frac{\hat{\theta}^{I}}{2}\, \gamma_5\right)^{-1}_{\rm oo}
  M_{\rm oe} 
\end{split}
\end{equation}
with
\begin{equation}
\hat{\theta}^{I} = \left(1 - \frac{\kappa}{\kappa_c}\right)\,
\bar{\theta}^{I}. 
\end{equation}
$M_{\rm eo}$ and $M_{\rm oe}$ are Wilson hopping matrices, which connect
even with odd and odd with even sites, respectively, and $T$ is the
clover matrix
\begin{equation}
T=\frac{i}{2} c_{SW}\, \kappa\, \sigma_{\mu\nu} F_{\mu\nu}(x).
\end{equation}

We apply mass preconditioning \`{a} la Hasenbusch~\cite{Hasenbusch} and split
the resulting action into three parts, each of which we put on separate time
scales~\cite{Sexton}. We use Omelyan's second order integrator~\cite{Omelyan}
to integrate Hamilton's equations of motion.

\section{Renormalization}

Let us now compute $Z_m^S Z_P$ for our action and coupling, so that we can 
compare the results to phenomenology later on. We demand that the renormalized
$VWI$ and axial vector ($AWI$) quark masses are equal, 
\begin{equation}
m_R = Z_m^S \, m = \frac{Z_A}{Z_P} \,\widetilde{m} = \widetilde{m}_R.
\end{equation}
That gives $Z_m^S Z_P = Z_A \, \widetilde{m}/m$. At the largest $\kappa$ value
(smallest quark mass) of~\cite{cppacs} we find $\widetilde{m}/m = 1.28(2)$. 
The renormalization constant $Z_A$ has been computed nonperturbatively
in~\cite{cppacs2} for a variety of couplings. Extrapolating the numbers to
$\beta=2.1$ gives $Z_A=0.78(1)$. Multiplying these two pieces of information
together, we obtain $Z_m^S Z_P = 1.00(5)$. That means $\theta_R =
\bar{\theta}$ to a good precision.  

Alternatively, we may compute $Z_m^S Z_P$ directly from $Z_m^S/Z_m^{NS}$ and
$Z_m^{NS} Z_P = Z_P/Z_S^{NS}$. As a comparison, the QCDSF Collaboration, using
nonperturbatively improved clover fermions and the plaquette gauge action,
finds at $\beta=5.4$~\cite{qcdsf} $Z_m^S/Z_m^{NS} = 1.25(5)$ and~\cite{qcdsf3}
$Z_P/Z_S^{NS} = 0.81(2)$. Altogether, this gives $Z_m^S Z_P = 1.01(5)$, in
agreement with the CP-PACS result, attesting clover fermions good chiral
properties. 

\section{Charge distribution and $\theta$ vacuum}

Before we compute the electric dipole moment now, let us look at the
distribution of topological charge and its dependence on $\theta$. Having
found that $\theta_R = \bar{\theta}$, the topological charge that follows from
the chirally rotated action (\ref{fa}) is the so-called fermionic charge
\begin{equation}
Q= \bar{m}\,{\rm Tr}\, \gamma_5\,M^{-1} ,
\label{qf}
\end{equation}
where $M = D + \bar{m}$ is the fermion matrix. The evaluation of (\ref{qf})
requires the computation of the $O(100)$ lowest-lying eigenvalues of $M$,
which is numerically expensive. It has been
demonstrated~\cite{Kovacs,Horsley1,Horsley2} that 
the fermionic charge and the so-called field theoretic charge, which is
computed from the field strength tensor by applying an appropriate number of
cooling sweeps and rounding the result to the nearest integer value, give
consistent results. In the following we shall employ the field theoretic
definition of the topological charge. It turns out that the results are rather
independent of the degree of cooling. In fact, the numbers stabilize already
after $O(10)$ cooling sweeps. The numbers quoted here refer to $O(100)$ cooling
sweeps. For a recent appraisal of the cooling method see~\cite{Ilgenfritz}. 

\begin{figure}[t]
\vspace*{-2cm}
\begin{center}
\epsfig{file=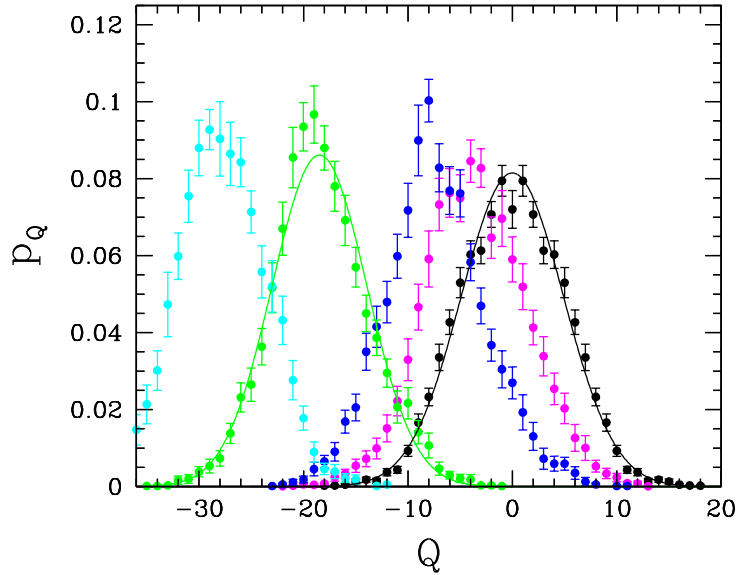,width=10cm,clip=}
\caption{The topological charge distribution for $\bar{\theta}^{I}=0$,
  $0.2$, $0.4$, $1.0$ and $1.5$, from right to left. To guide the eye, the
  distribution is compared to a Gaussian fit at $\bar{\theta}^{I}=0$
  and $1.0$.}  
\end{center}
\end{figure}

\begin{table}[b]
\begin{center}
\begin{tabular}{c|c|c} $\bar{\theta}^{I}$ & $\langle Q\rangle$ & $\langle
Q^2\rangle_c$ \\ \hline
0\phantom{.0} & \phantom{0}-0.06(31) & 24.9(14) \\
0.2 & \phantom{0}-3.52(46) &  24.1(15) \\
0.4 & \phantom{0}-7.35(36) &  22.7(17) \\
1.0 & -18.38(30) &  21.7(15) \\
1.5 & -27.84(37) &  18.1(13)
\end{tabular}
\caption{The average topological charge and charge squared for our values
  of $\bar{\theta}^{I}$.}  
\end{center}
\label{avcharge}
\end{table}

In Fig.~1 we show the charge distribution for our five different values of
$\bar{\theta}^{I}$. To identify the shape of the distribution, and to see how
it changes with increasing value of $\bar{\theta}^{I}$, we compare our data to
a Gaussian fit at $\bar{\theta}^{I} =0$ and $1.0$. A Gaussian distribution is
the most common distribution function for independent, randomly generated
variables. 

In Table~I we present the resulting average topological charge $\langle
Q\rangle$ and charge squared, 
\begin{equation}
\langle Q^2\rangle_c \equiv \left\langle (Q - \langle Q\rangle)^2 \right\rangle
= \langle Q^2\rangle - \langle Q\rangle^2, 
\end{equation}
which we plot in Figs.~2 and 3. As with any distribution, besides the
distributions mean, its skewness and kurtosis coefficients should be calculated
in order to determine the type of distribution. We do so in Figs.~4 and 5,
where we plot the skewness $S$,
\begin{equation}
S= \frac{\langle Q^3\rangle_c}{\langle Q^2\rangle_c}\, , \quad \langle
Q^3\rangle_c \equiv \left\langle (Q - \langle Q\rangle)^3 \right\rangle,
\end{equation}
and kurtosis $K$,
\begin{equation}
K= \frac{\langle Q^4\rangle_c}{\langle Q^2\rangle_c}\, , \quad \langle
Q^4\rangle_c \equiv \left\langle (Q - \langle Q\rangle)^4 \right\rangle -
3 \left\langle (Q - \langle Q\rangle)^2 \right\rangle^2.
\end{equation}
Note that $S$ and $K$ have been normalized to $\langle Q^2\rangle_c$,
different from the mathematical literature, so that the volume dependence
cancels out. 

Let us first look at the charge distributions in Fig.~1. At
$\bar{\theta}^{I}=0$ the lattice data show a higher probability than a
Gaussian distributed charge for intermediate values of $|Q|$ and a slightly
thinner tail, while at larger values of $\bar{\theta}^{I}$ the distributions
appears to show a sharper peak and fatter tail. On top of that, the right tail
becomes longer compared to the left one with increasing value of
$\bar{\theta}^{I}$.  

This behavior is reflected in the skewness and kurtosis coefficients shown in
Figs.~4 and 5. Skewness is a measure of the degree of asymmetry of a
distribution. If the right tail of the distribution is more pronounced than
the left one, the distribution is said to have positive skewness, which is
what we observe at $\bar{\theta}^{I}>0$. If the reverse is true, it has
negative skewness. Kurtosis is the degree of peakedness of a distribution. A
distribution with positive kurtosis is called leptokurtic and has an acute
peak around its mean. Examples of leptokurtic distributions include the
Laplace distribution. A distribution with negative kurtosis is called
platykurtic and has a smaller peak around its mean and a lower probability
than a Gaussian distribution at extreme values. Such distributions are termed
sub-Gaussian.~\footnote{A distribution $P(x)=\exp[-f(x^2)]$ is called
  sub-Gaussian if $f^\prime(x^2)$ is strictly increasing on $[0,\infty)$.} 
The kurtosis starts out negative at $\bar{\theta}^{I}=0$ and rises almost
linearly to reach positive values at $\bar{\theta}^{I}\gtrsim 0.2$. For recent
quenched results see~\cite{DelDebbio,Durr,Giusti}. 

\clearpage
\begin{figure}
\begin{center}
\vspace*{-2.0cm}
\epsfig{file=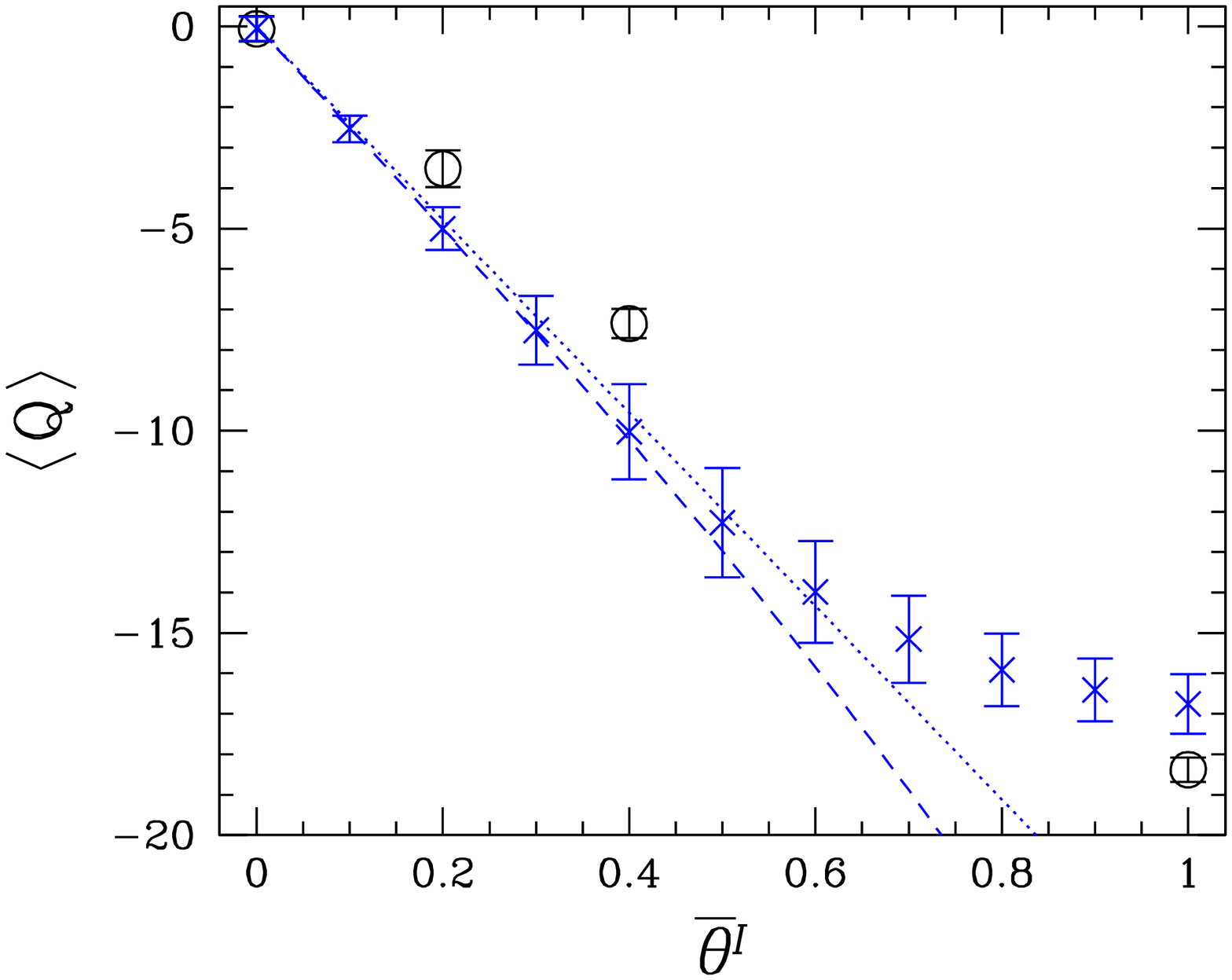,width=9.75cm,clip=}
\caption{The average charge (${\Circle}$) compared to the reweighted numbers
  ($\times$) and to the predictions of the Gaussian distribution (dotted line)
  and the dilute instanton gas (\ref{qgas}) (dashed line).}  
\end{center}
\label{Qfig}
\end{figure}

\begin{figure}
\begin{center}
\vspace*{-2.0cm}
\epsfig{file=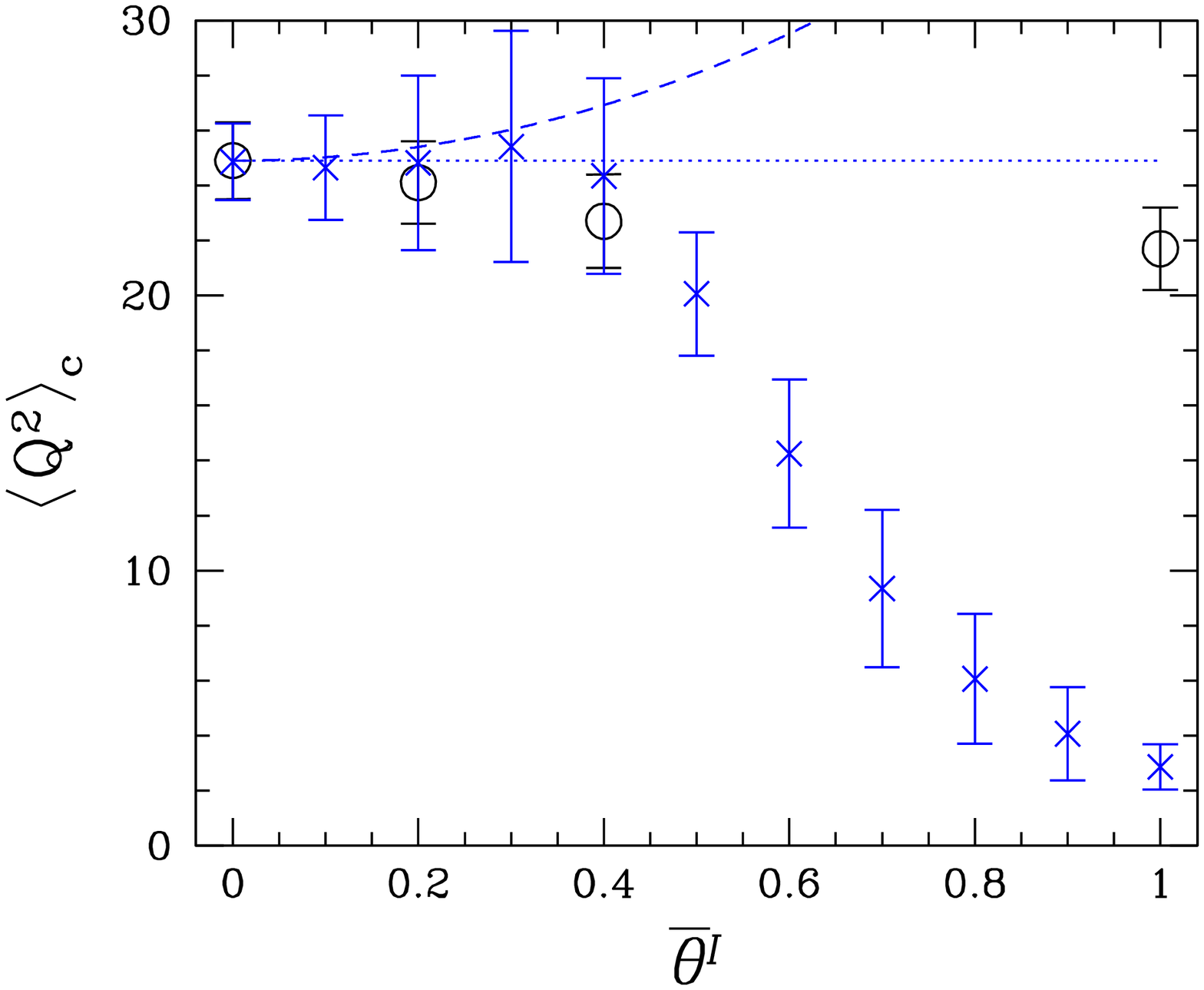,width=9.75cm,clip=}
\caption{The average charge squared (${\Circle}$) compared to the reweighted
  numbers ($\times$) and to the predictions of the Gaussian distribution
  (dotted line) and the dilute instanton gas (\ref{qgas}) (dashed line).}
\end{center}
\label{Qfig2}
\end{figure}

\clearpage
\begin{figure}
\vspace*{-2.0cm}
\begin{center}
\epsfig{file=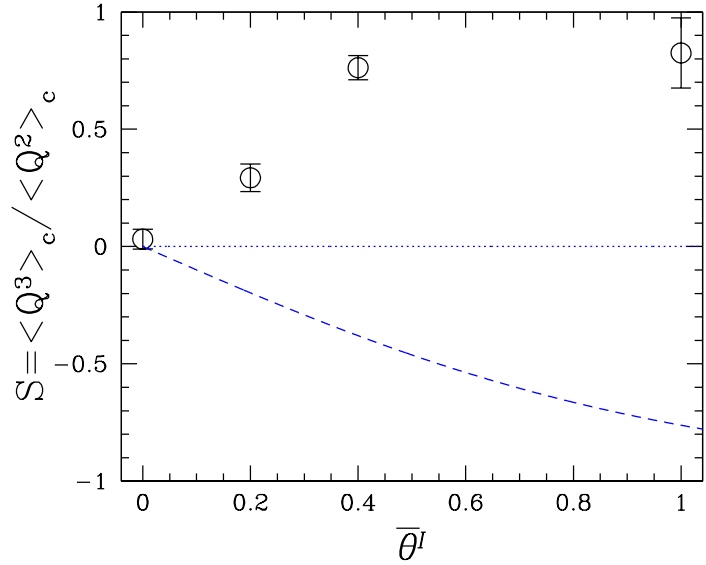,width=9.75cm,clip=}
\caption{The skewness compared to the predictions of the Gaussian distribution
  (dotted line) and the dilute instanton gas (dashed line). The errors shown
  are the naive ones, as our sample of large charges $|Q-\langle
  Q\rangle|$ is too small to allow for a proper jackknive analysis, and may
  be underestimated.}   
%The result for $\bar{\theta}^{I}=1.5$ is not shown because of large errors.} 
\end{center}
\end{figure}

\begin{figure}[t]
\vspace*{-2.0cm}
\begin{center}
\epsfig{file=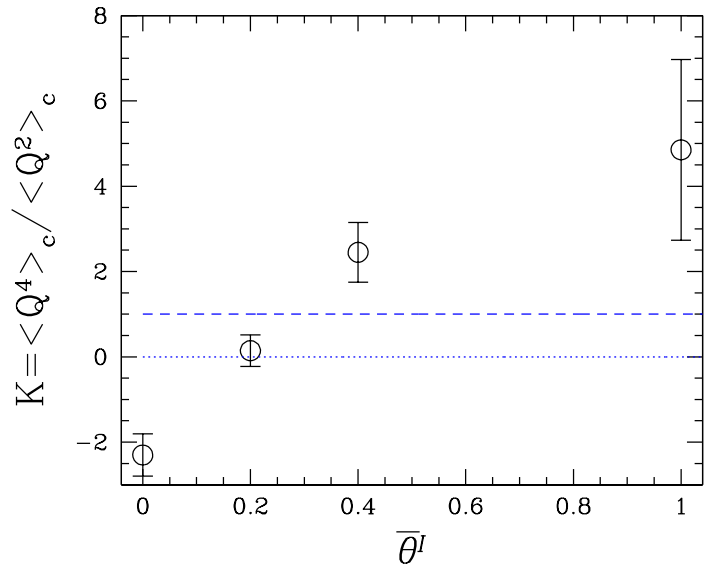,width=9.75cm,clip=}
\caption{The Kurtosis, together with its naive error, compared to the
  predictions of the Gaussian distribution 
  (dotted line) and the dilute instanton gas (dashed line).}
%The result for $\bar{\theta}^{I}=1.5$ is not shown because of large errors.} 
\end{center}
\end{figure}

\clearpage
A popular model of the QCD vacuum is the dilute instanton gas~\cite{cdg},
whose charge distribution is given by a convolution of separate Poisson
distributions for instantons and anti-instantons. Its free energy per
spacetime volume V is 
\begin{equation}
F(\theta)= \chi_t \, (1-\cos \theta), \quad F(\theta)  = - \frac{1}{V} \,
\ln \, Z(\theta), 
\end{equation}
where 
\begin{equation}
\chi_t=\left.\frac{\langle Q^2\rangle}{V}\right|_{\theta=0} .
\end{equation}
That leads to~\cite{klssw} 
\begin{equation}
%\begin{split}
\langle Q^n\rangle_c = - i^n\, V \frac{\partial^n F(\theta)}{\partial
  \theta^n} ,
%\end{split}
\end{equation}
which at imaginary $\theta$ gives
\begin{equation}
\begin{split}
\langle Q\rangle &= - V \chi_t\, \sinh \theta^{I}, \\
\langle Q^2\rangle_c & = V \chi_t\, \cosh \theta^{I}, \\
\langle Q^3\rangle_c & = - V \chi_t\, \sinh \theta^{I}, \\
\langle Q^4\rangle_c & = V \chi_t\, \cosh \theta^{I}.
\end{split}
\label{qgas}
\end{equation}

In Figs.~2-5 we compare the lattice data for $\langle Q\rangle$, $\langle
Q^2\rangle_c$, $S$ and $K$ with the predictions of the Gaussian distribution
and the dilute instanton gas. While $\langle Q\rangle$ and $\langle
Q^2\rangle_c$ are in reasonable agreement with the results of the Gaussian
and Poisson distribution for smaller values of $\bar{\theta}^{I}$, the higher
cumulants $S$ and $K$ show a far different trend over the entire range of
$\bar{\theta}^{I}$ than the predictions of these simple models.

Reweighting appears to be the accepted method for simulations at nonvanishing
vacuum angle $\theta$ and chemical potential $\mu$. Having results of a direct
simulation at nonvanishing value of $\theta$ at hand, it is instructive to
test how reliable the method actually is, given the fact that the simulations
are necessarily restricted to a finite volume with limited absolute value of
the topological charge.

The reweighted charges are given by
\begin{equation}
\langle Q^n \rangle = \frac{1}{Z(\theta)} \sum_Q Q^n \,P_Q \,{\rm
  e}^{-i\,\theta\, Q}, \quad \sum_Q P_Q = 1,
\end{equation}
where $P_Q$ denotes the probability of finding a configuration of
charge $Q$ in the ensemble of configurations, and
\begin{equation}
Z(\theta) = \sum_Q P_Q \,{\rm  e}^{-i\,\theta\, Q}.
\end{equation}
At imaginary $\theta = -i\,\theta^{I}$ this becomes
\begin{equation}
\langle Q^n \rangle = \frac{1}{Z(\theta)} \sum_Q Q^n \,P_Q \,{\rm
  e}^{- \theta^{I}\, Q}, \quad Z(\theta) = \sum_Q P_Q \,{\rm
  e}^{- \theta^{I}\, Q}. 
\end{equation}

In Figs.~2 and 3 we compare $\langle Q\rangle$ and $\langle Q^2\rangle_c$ with
the reweighted numbers, where we have converted $\theta$ to $\bar{\theta}$
using (\ref{bar}). On the quantitative level, reweighting is not able to
describe the data for $\bar{\theta}^{I}\geq 0.4$. The reason is that the
reweighted charge distributions have largely the same form as the initial
$\bar{\theta}^{I}=0$ distribution, but are merely shifted towards negative $Q$
values, while the shape of the true charge distributions changes significantly
with increasing value of $\bar{\theta}^{I}$. We may expect to find better
agreement on larger volumes, provided $Z(\theta)$ is analytic in $\theta$. 
%We note that this comparison should not depend on the method used for
%calculating the topological charge, since it is done the same way in both
%cases.   

\section{Nucleon form factors at $\mathbf{\theta \neq 0}$}

At nonvanishing $\theta$ the electromagnetic current between nucleon states
can be decomposed in Euclidean space into
\begin{equation}
\langle p^\prime,s^\prime|J_\mu|p,s\rangle =
\bar{u}_\theta(\vec{p}^{\,\prime},s^\prime)\, \mathcal{J}_\mu\,
u_\theta(\vec{p},s),
\label{me}
\end{equation}
with
\begin{equation}
\mathcal{J}_\mu = \gamma_\mu F_1^\theta(q^2) + \sigma_{\mu\nu} q_\nu
\frac{F_2^\theta(q^2)}{2m_N^\theta} +  \left[ (\gamma q\, q_\mu -
  \gamma_\mu \, q^2) \, \gamma_5 \, F_A^\theta(q^2) + \sigma_{\mu\nu} q_\nu \,
  \gamma_5 \frac{F_3^\theta(q^2)}{2m_N^\theta}\right],
\label{current}
\end{equation}
where $q=p^\prime - p$. The form factors and nucleon mass will generally
depend on $\theta$ with $F_{\cdots}^{\theta=0} = F_{\cdots}$ and
$m_N^{\theta=0} = m_N$. The Dirac spinors are modified by a phase in the
$\theta$ vacuum,
\begin{equation}
\begin{split}
u_\theta(\vec{p},s) &= {\rm e}^{i\alpha(\theta)\gamma_5}\, u(\vec{p},s),\\
\bar{u}_\theta(\vec{p},s) &= \bar{u}(\vec{p},s)\, {\rm
  e}^{i\alpha(\theta)\gamma_5}, 
\end{split}
\end{equation}
so that the standard spinor relation is modified to
\begin{equation}
\sum_{s^\prime,s} u_\theta(\vec{p},s^\prime) \bar{u}_\theta(\vec{p},s) = {\rm
  e}^{i\alpha(\theta)\gamma_5} \left(\frac{-i \gamma p + m_N^\theta}{2
  E_N^\theta}\right) {\rm  e}^{i\alpha(\theta)\gamma_5}.
\end{equation}
As we are primarily interested in the electric dipole moment in the limit
$\theta \rightarrow 0$, it is sufficient to consider the lowest order
expansion only. Hence, we may write 
\begin{equation}
\alpha(\theta) = \alpha^\prime \, \theta + O(\theta^3).
\label{phase}
\end{equation}
For our choice of $\theta$ (\ref{choice}), this then becomes to lowest order
in $\theta$
\begin{equation}
\sum_{s^\prime,s} u_\theta(\vec{p},s^\prime) \bar{u}_\theta(\vec{p},s) =
\frac{-i \gamma p + m_N (1+2\alpha^\prime \bar{\theta}^{I} \gamma_5)}{2 E_N}.
\end{equation}
Note that in Euclidean space $q^2 = - (E^\prime - E)^2 + (\vec{p}^{\,\prime} -
\vec{p})^2$ and $\gamma p = i E \gamma_4 + \vec{\gamma}\vec{p}$.

We denote the two-point function of a nucleon of momentum $\vec{p}$ in the
theta vacuum by $G_{NN}^\theta(t,\vec{p})$. The phase factor $\alpha^\prime$
of (\ref{phase}) can be obtained from the ratio of two-point functions
\begin{equation}
\begin{split}
{\rm Tr}\, [G_{NN}^\theta(t;0) \Gamma_4] &\simeq \frac{1}{2} |Z_N|^2\, {\rm
  e}^{-m_N t}, \\
{\rm Tr}\, [G_{NN}^\theta(t;0) \Gamma_4 \gamma_5] &\simeq - \alpha^\prime
  \bar{\theta}^{I}\, \frac{1}{2} |Z_N|^2\, {\rm
  e}^{-m_N t},
\end{split}
\end{equation}
where $\Gamma_4 = (1+\gamma_4)/2$. In Fig.~6 we show
\begin{equation}
R(t) = \frac{{\rm Tr}\, [G_{NN}^\theta(t,0) \Gamma_4 \gamma_5]}{{\rm Tr}\,
  [G_{NN}^\theta(t;0) \Gamma_4]} \simeq - \alpha^\prime \bar{\theta}^{I}
\label{ratio}
\end{equation}
for $\bar{\theta}^{I}=0.4$. Fitting to a constant, we find
\begin{equation}
\begin{tabular}{c|c}
$\bar{\theta}^{I}$ & $\alpha^\prime \bar{\theta}^{I}$  \\ \hline
0.2 & 0.048(3) \\
0.4 & 0.081(4)
\end{tabular}
\end{equation}
\begin{figure}[t]
\vspace*{0.25cm}
\begin{center}
\epsfig{file=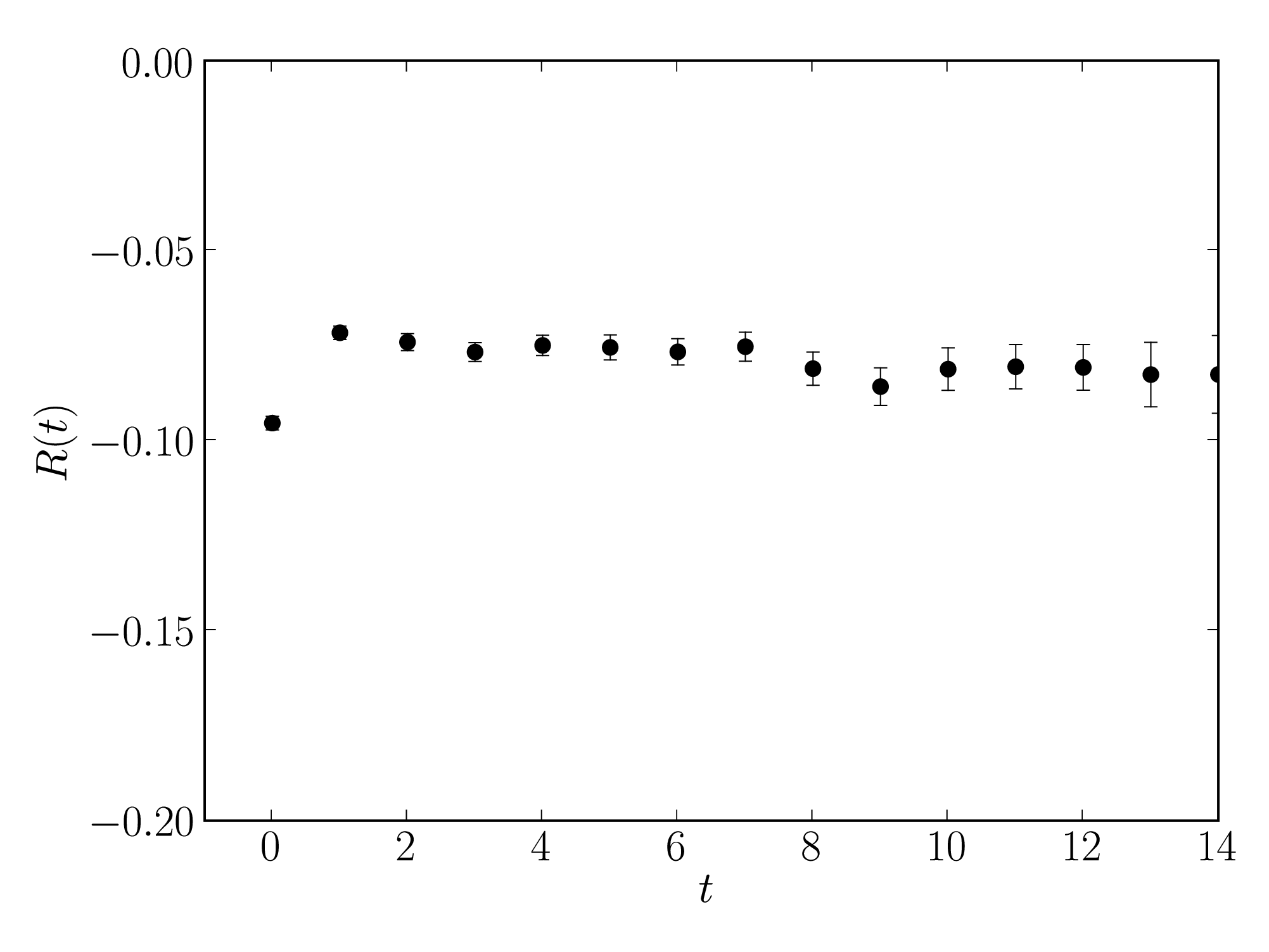,width=10cm,clip=}
\caption{The ratio $R(t$) given in (\ref{ratio}) for $\bar{\theta}^{I} = 0.4$.}
\end{center}
\end{figure}
The form factor $F_3(q^2)$, which is needed for the determination of the
nucleon electric dipole moment, can be extracted from the ratio of three-point
and two-point functions
\begin{equation}
\begin{split}
R_\mu(t^\prime,t;\vec{p}^{\,\prime},\vec{p}) &= \frac{G_{NJ_\mu
    N}^{\theta\, \Gamma}(t^\prime,t;\vec{p}^{\,\prime},\vec{p})}{{\rm Tr}\,
    [G_{NN}^\theta(t^\prime;\vec{p}^{\,\prime})\Gamma_4]} \\[0.5em]
    &\times \left\{\frac{{\rm
    Tr}\,[G_{NN}^\theta(t;\vec{p}^{\,\prime})\Gamma_4]\,  
    {\rm Tr}\,[G_{NN}^\theta(t^\prime;\vec{p}^{\,\prime})\Gamma_4]\,
    {\rm Tr}\,[G_{NN}^\theta(t^\prime-t;\vec{p})\Gamma_4]}
    {{\rm Tr}\,[G_{NN}^\theta(t;\vec{p})\Gamma_4]\, 
    {\rm Tr}\,[G_{NN}^\theta(t^\prime;\vec{p})\Gamma_4]\,
    {\rm
    Tr}\,[G_{NN}^\theta(t^\prime-t;\vec{p}^{\,\prime})\Gamma_4]}\right\}^{1/2}
    \\[0.75em] 
&= \sqrt{\frac{E^{\theta\,\prime} \,E^\theta}
   {(E^{\theta\,\prime}+m_N^\theta) \,
    (E^\theta+m_N^\theta)}} \, F(\Gamma,\mathcal{J}_\mu),
\end{split}
\label{rw}
\end{equation}
where $G_{NJ_\mu N}^{\theta\, \Gamma}((t^\prime,t;\vec{p}^{\,\prime},\vec{p})$
is the three-point function, with $t^\prime$ being the time location of
the nucleon sink and $t$ the time location of the current insertion, and the
function $F(\Gamma,\mathcal{J}_\mu)$ is
\begin{equation}
\begin{split}
F(\Gamma,\mathcal{J}_\mu) = \frac{1}{4}\, {\rm Tr}\, \Gamma &\left[{\rm
    e}^{i\alpha(\theta)\gamma_5} \frac{E^{\theta\,\prime}\gamma_4
    -i\vec{\gamma}\vec{p}^{\,\prime} + m_N^\theta}{E^{\theta\,\prime}}\,
    {\rm  e}^{i\alpha(\theta)\gamma_5}\right]\\[0.5em]
&\times \mathcal{J}_\mu \left[{\rm
    e}^{i\alpha(\theta)\gamma_5} \frac{E^{\theta}\gamma_4
    -i\vec{\gamma}\vec{p} + m_N^\theta}{E^{\theta}}\,
    {\rm  e}^{i\alpha(\theta)\gamma_5}\right]
\end{split}
\end{equation}
with $\mathcal{J}_\mu$ given in (\ref{current}). The three-point functions are
calculated for various choices of nucleon polarization, $\Gamma = \Gamma_4$,
$i\Gamma_4\gamma_5\gamma_1$ and $i\Gamma_4\gamma_5\gamma_2$. The calculation
follows the program of the QCDSF Collaboration~\cite{qcdsf4} for computing
nucleon three-point functions. We neglect finite volume corrections to two-
and three-point functions~\cite{Liu}.
For $J_\mu$ we take the local vector current, which needs to be
renormalized. We compute the corresponding renormalization constant
$Z_V$~\cite{qcdsf2} from the proton form factor $F_1(0)$ at zero momentum
transfer. 

With conventional (periodic) boundary conditions momenta are quantized in
units of $2\pi/L$, where $L$ is the spatial extent of the lattice. For the
lattices used in the current simulation, this means that the smallest
nonvanishing momentum available is $\approx 700 \, \mbox{MeV}$. Since $F_3$
can only be computed at $q^2 \neq 0$, we need to extrapolate to $q^2=0$ to
find $d_N$.
The momentum resolution of hadron observables can be significantly improved by
varying the boundary conditions. It was demonstrated~\cite{Sachrajda} that for
processes without final state interactions, such as the form factors studied
in this paper, it is sufficient to apply twisted boundary conditions to the
valence quarks only. 

In our study we use partially twisted boundary conditions, {\it i.e.}\
combining gauge field configurations generated with sea quarks with periodic
spatial boundary conditions with valence quarks with twisted boundary
conditions. The boundary conditions of the valence quarks attached to the
electromagnetic current are 
\begin{equation}
\psi(x_k + L) = {\rm e}^{i\,\alpha_k} \, \psi(x_k), \quad k=1, 2, 3.
\end{equation} 
By varying $\vec{\alpha}$ we
can tune the momenta of the nucleon continuously. We have chosen the following
set of twist angles
\begin{equation}
\begin{split}
\vec{\alpha} &= \frac{2\pi}{L} \, (0,0,0) \\
\vec{\alpha} &= \frac{2\pi}{L} \, (0.36,0,0) \\
\vec{\alpha} &= \frac{2\pi}{L} \, (0.36,0.36,0) \\
\vec{\alpha} &= \frac{2\pi}{L} \, (0.36,0.36,0.36) 
\end{split}
\end{equation}
The dispersion relation for the nucleon then reads
\begin{equation}
E=\sqrt{m_N^2 + (\vec{p}+\vec{\alpha})^2} .
\end{equation}

We define the electric dipole moment as
\begin{equation}
d_N^\theta = \frac{e\,F_3^\theta(0)}{2m_N^\theta}.
\end{equation}
In Fig.~7 we show our results for $F_3$ for proton and neutron on
configutaions with $\bar{\theta}^{I}=0.2$ (top) and $\bar{\theta}^{I}=0.4$
(bottom). We find a very 
clean signal for the neutron form factor. The signal for the proton, on the
other hand, is somewhat more noisy. The reason is that in this case one has to
subtract $F_1^p(0)$ from the matrix element (\ref{current}), which is by far
the largest contribution.  

To obtain the
result at $q^2=0$, we first attempt a fit using a dipole ansatz
\begin{equation}
F_3^\theta(q^2) = \frac{F_3^\theta(0)}{(1+q^2/M^2)^2},
\end{equation}
which is indicated by the solid lines. At $q^2=0$ we find
\begin{equation}
\begin{tabular}{c|c|c} 
$\bar{\theta}^{I}$ & $F_3^p(0)/(2m_N)$ & $F_3^n(0)/(2m_N)$\\ \hline
0.2 & $\phantom{-}0.158(33)$ & $-0.108(17)$ \\
0.4 & $\phantom{-}0.256(25)$ & $-0.193(12)$ 
\end{tabular}
\label{dp}
\end{equation}

Alternatively, if we assume that $F_3$ and $F_1^p$ have similar $q^2$
behavior, then by forming the ratio
\begin{equation}
\frac{F_3^\theta(q^2)}{F_1^{\theta\, p}(q^2)},
\end{equation}

\clearpage
\begin{figure}[t!]
\vspace*{0.5cm}
\begin{center}
\epsfig{file=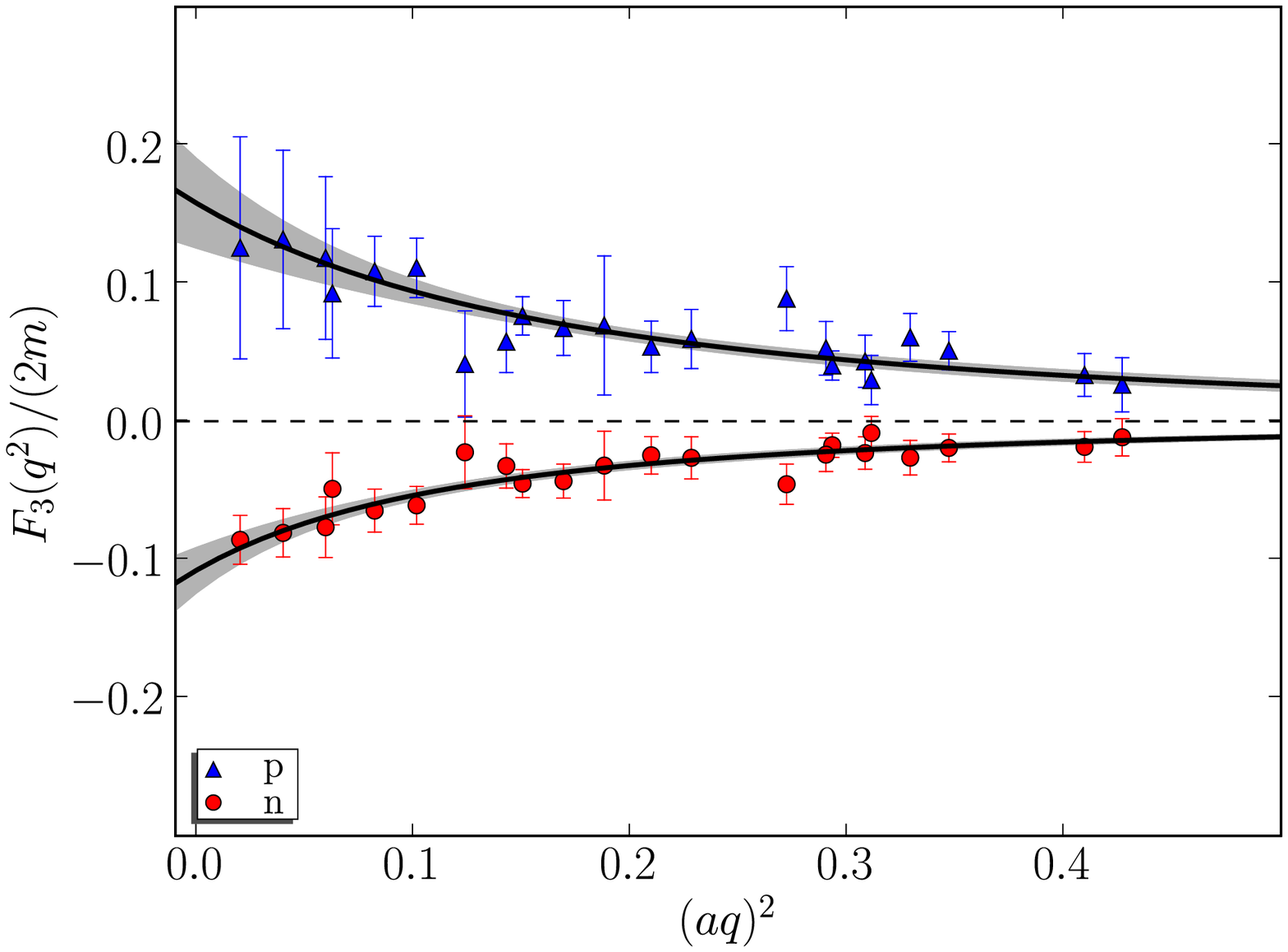,width=11cm,clip=}\\[1cm]
\epsfig{file=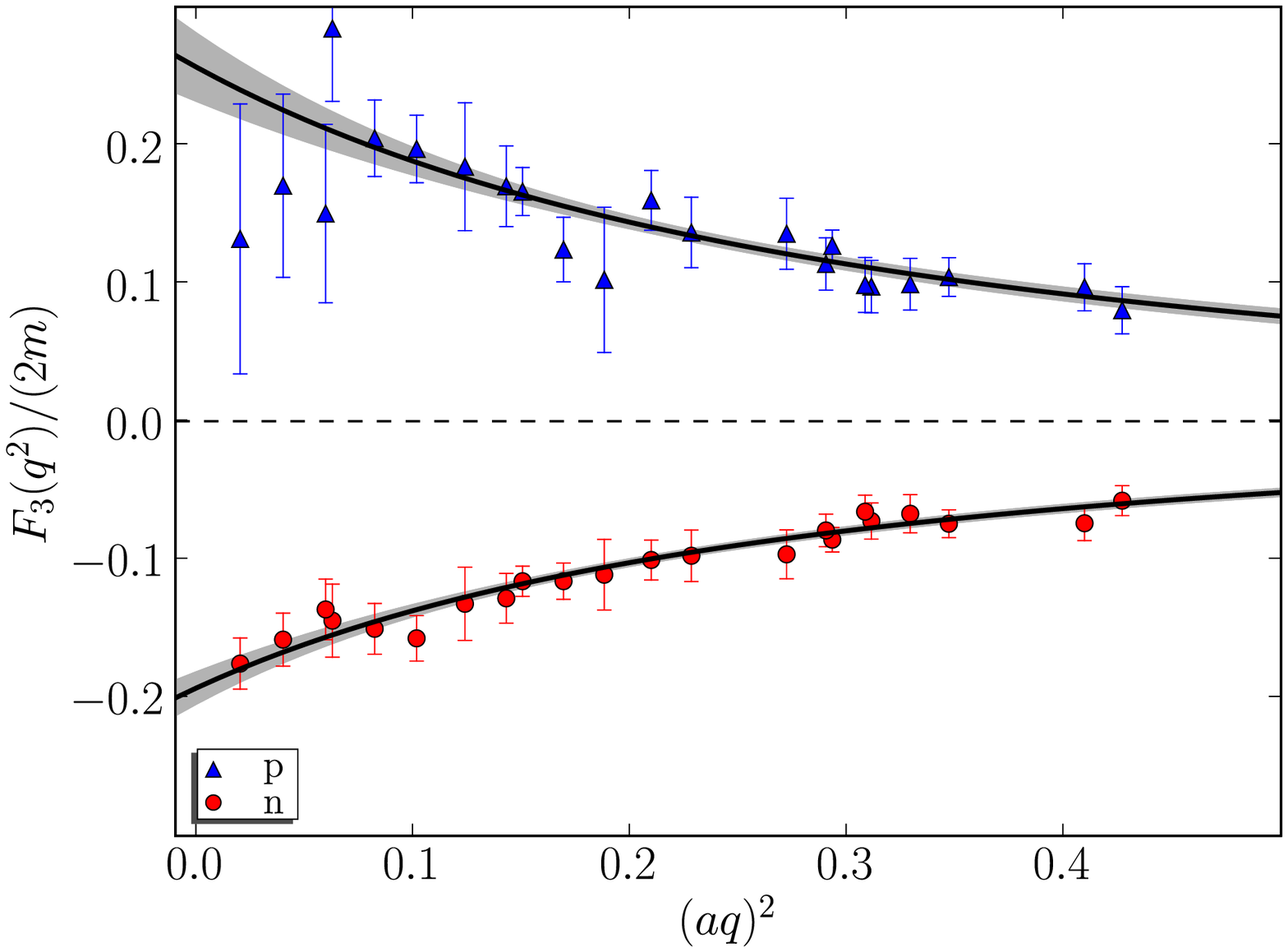,width=11cm,clip=}
\vspace*{0.4cm}
\caption{The form factor $F_3(q^2)$ for proton and neutron, together with a
  dipole fit, for $\bar{\theta}^{I}=0.2$ (top) and $\bar{\theta}^{I}=0.4$
  (bottom), respectively.}
\end{center}
\end{figure}

\clearpage
\begin{figure}[t!]
\vspace*{0.5cm}
\begin{center}
\epsfig{file=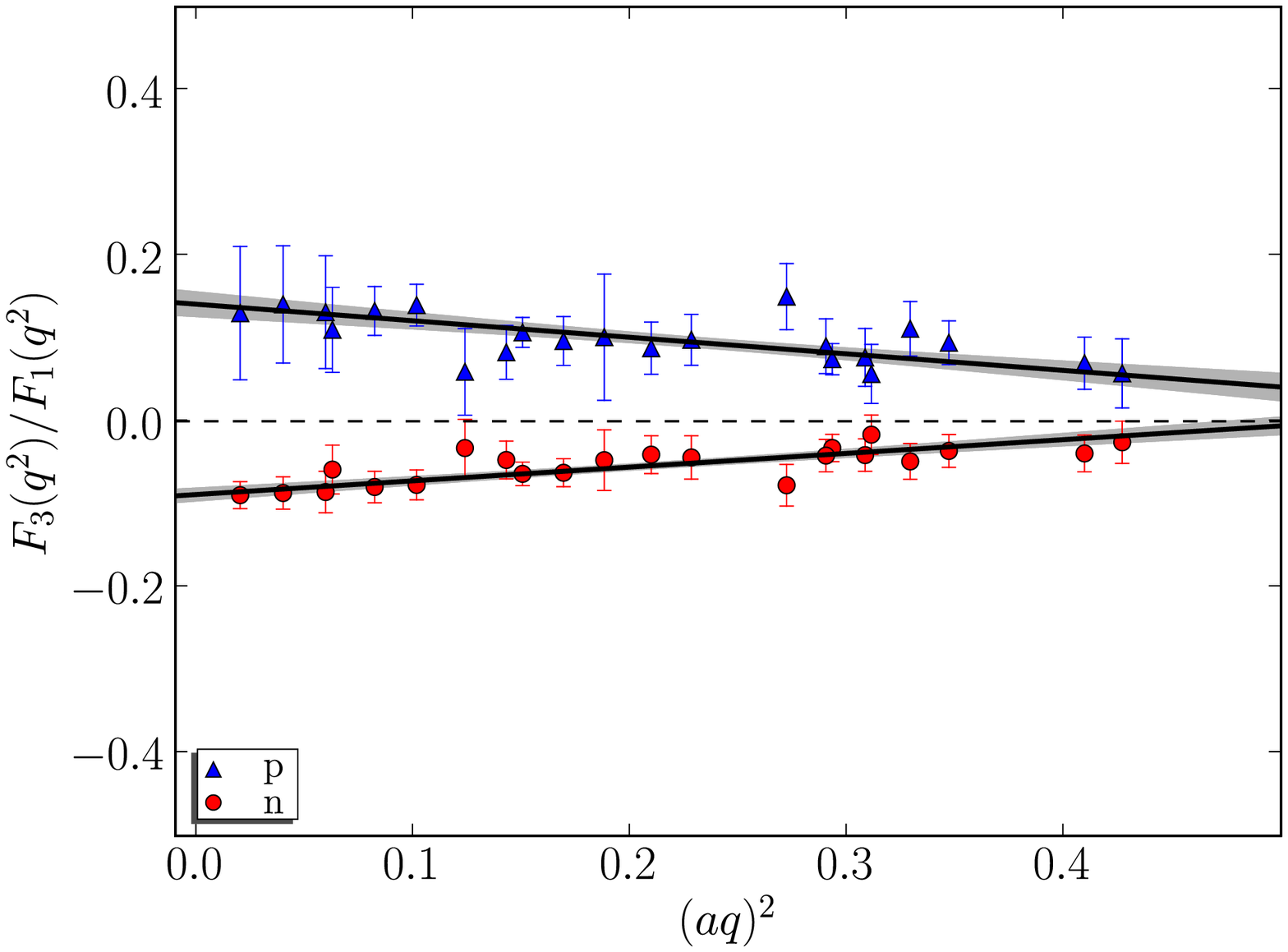,width=11cm,clip=}\\[1cm]
\epsfig{file=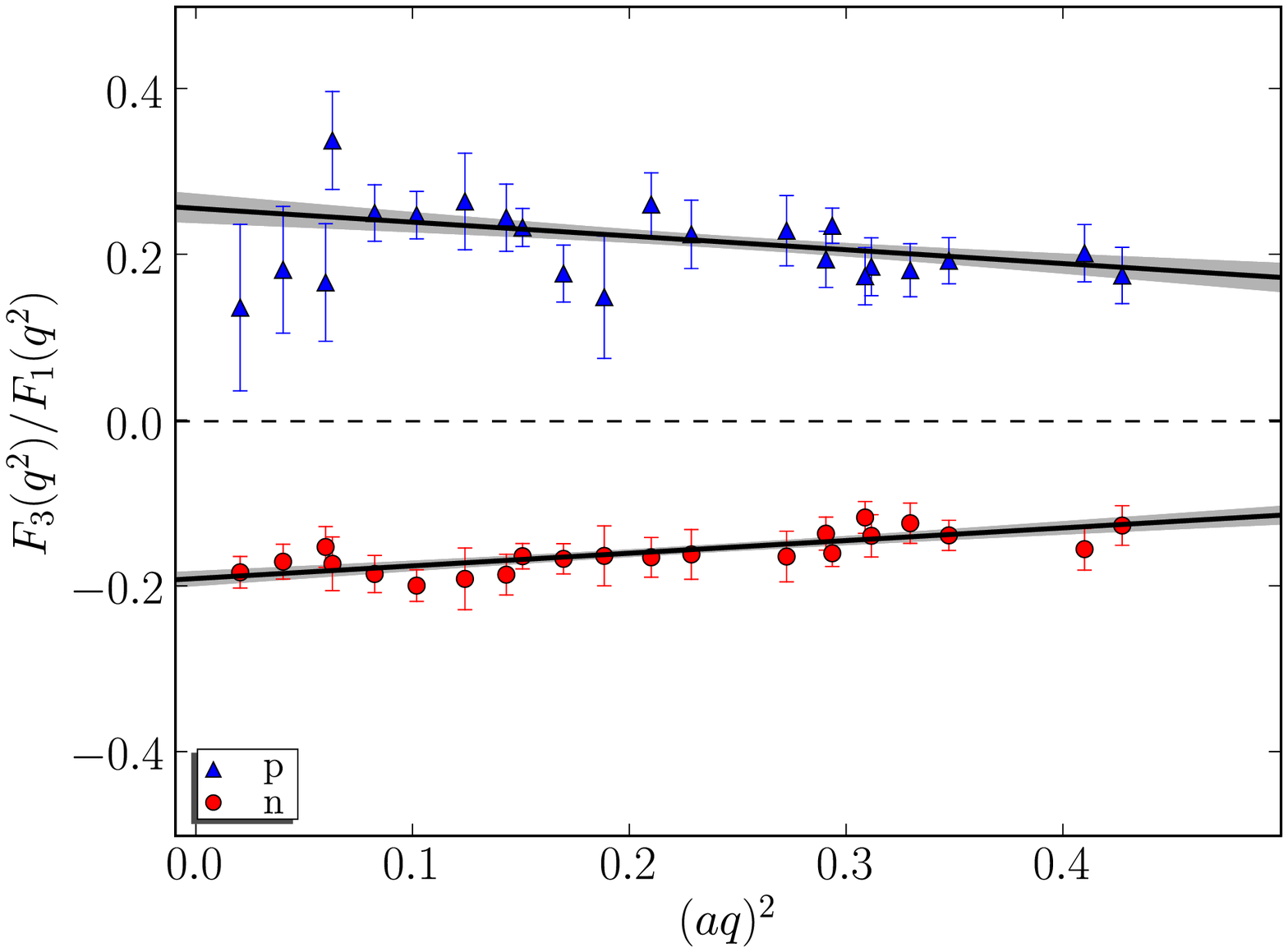,width=11cm,clip=}
\vspace*{0.4cm}
\caption{The form factor ratio $F_3(q^2)/F_1(q^2)$ for proton and neutron, for
  $\bar{\theta}^{I}=0.2$ (top) and $\bar{\theta}^{I}=0.4$ (bottom),
  respectively.} 
\end{center}
\end{figure}

\clearpage
\noindent
the renormalization constant $Z_V$ cancels, and we may hope to see a constant
behavior as a function of $q^2$. In Fig.~8 we show this ratio. Again, we find
a very clean signal for the neutron. After performing a linear extrapolation
to $q^2=0$, we obtain 
\begin{equation}
\begin{tabular}{c|c|c} 
$\bar{\theta}^{I}$ & $F_3^p(0)/(2m_N)$ & $F_3^n(0)/(2m_N)$\\ \hline
0.2 & $\phantom{-}0.141(16)$ & $-0.088(8)\phantom{0}$ \\
0.4 & $\phantom{-}0.257(18)$ & $-0.190(9)\phantom{0}$ 
\end{tabular}
\label{ra}
\end{equation}

\section{Electric dipole moment}

\begin{figure}[t]
\vspace*{-2cm}
\begin{center}
\epsfig{file=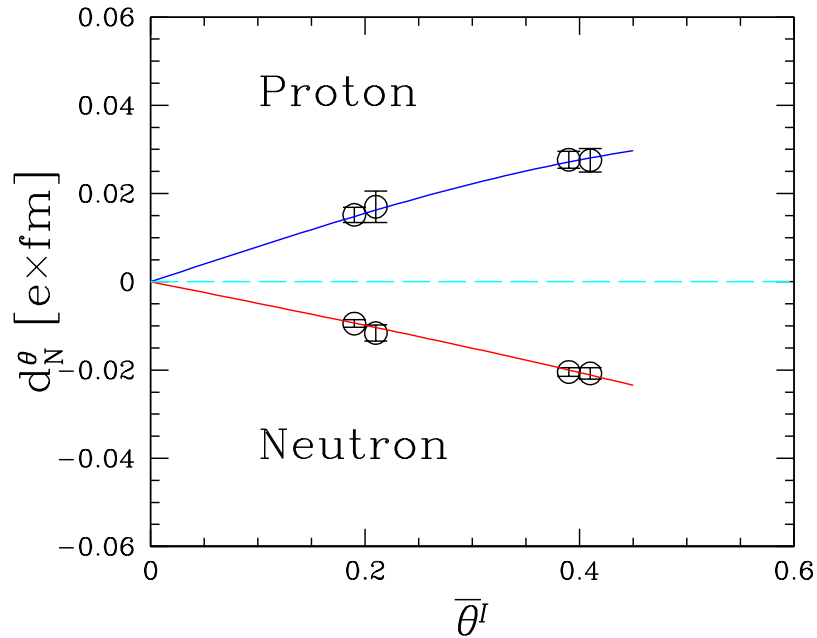,width=11cm,clip=}
\caption{The electric dipole moment $d_N^\theta$ for proton and neutron,
  together with a linear plus cubic fit. The data points are horizontally
  displaced for better legibility.}
\end{center}
\end{figure}

Both values of $F_3(0)$, (\ref{dp}) and (\ref{ra}), are consistent with each
other within the error bars. In Fig.~9 we show the results together with a fit
of the form
\begin{equation}
d_N^\theta = \frac{\partial d_N^\theta}{\partial \bar{\theta}^{I}}\,
\bar{\theta}^{I} + c\,\bar{\theta}^{I\; 3},
\end{equation}
as we are mainly interested in the derivative $\partial d_N^\theta/\partial
\bar{\theta}^{I}$. The fit gives at $\bar{\theta}^{I} =0$
\begin{equation}
\begin{split}
\frac{\partial d_N^\theta}{\partial \bar{\theta}^{I}} &=
\phantom{-}0.080(10)\;\; [e \times \mbox{fm}]  \quad \mbox{Proton}, \\[0.35em]
\frac{\partial d_N^\theta}{\partial \bar{\theta}^{I}} &=
-0.049(5)\phantom{0}\;\; [e \times \mbox{fm}]  \quad \mbox{Neutron}. 
\end{split}
\end{equation}
Combining the upper experimental bound on the electric dipole moment of the
neutron (\ref{ub}) with our result for $\displaystyle \partial
d_N^\theta/\partial \bar{\theta}^{I}$, we may derive an upper bound on the
vacuum angle $\theta$. Taking our results at face value, we find
\begin{equation}
|\theta| < 6 \times 10^{-12}.
\label{bound}
\end{equation}
It should be noted, however, that we are working at unphysically large quark
mass yet, so that this result has limited phenomenological significance.
%For comparison, QCD sum rules give~\cite{Pospelov} $\displaystyle |\theta| < 3
%\times 10^{-10}$.  

\section{Conclusion and outlook}

We have performed simulations of QCD with $N_f=2$ flavors of dynamical quarks
at imaginary vacuum angle $\theta$. It is the first time this has been done in
full QCD. The use of partially twisted boundary conditions has allowed us to
compute the proton and neutron form factor $F_3(q^2)$ with high precision over
the entire range of momenta down to $(aq)^2 \approx 0.02$, which greatly
facilitated the extrapolation to $q^2=0$.  

A further improvement of our calculation is that it does not require
reweighting of the three-point functions (\ref{rw}) with the topological
charge. Barring the fact that the lattice definition of topological charge is
ambiguous, to some extent, reweighting does not describe the charge
distribution accurately beyond $\bar{\theta}^{I} > 0.2$, which casts some
doubts on the method, if taken at face value. 

Having demonstrated the benefit of simulations at imaginary $\theta$, the next
step is to extend the calculations to more realistic quark masses and larger
lattices. The idea then is to make contact to the predictions of chiral
perturbation theory, which allow for the extrapolation of the dipole moment
from finite to infinite volume and to the physical quark 
mass~\cite{Borasoy,Savage}. Chiral perturbation theory also predicts the
$\theta$ dependence of physical quantities, such as hadron masses. For the
pion mass one finds~\cite{Brower}
\begin{equation}
m_\pi^2(\theta)=m_\pi^2(0)\,\cos(\theta/N_f).
\label{mpitheta}
\end{equation}
Though our quark mass is rather heavy, it is tempting to compare our results
with (\ref{mpitheta}). This is done in Fig.~10. The variation of $m_\pi$ with
$\theta$ is found to be significant. Our results do not confirm the
predictions of chiral perturbation theory. It will be interesting to see if
this behavior persists at smaller quark masses. 

\begin{figure}[t]
\vspace*{-2cm}
\begin{center}
\epsfig{file=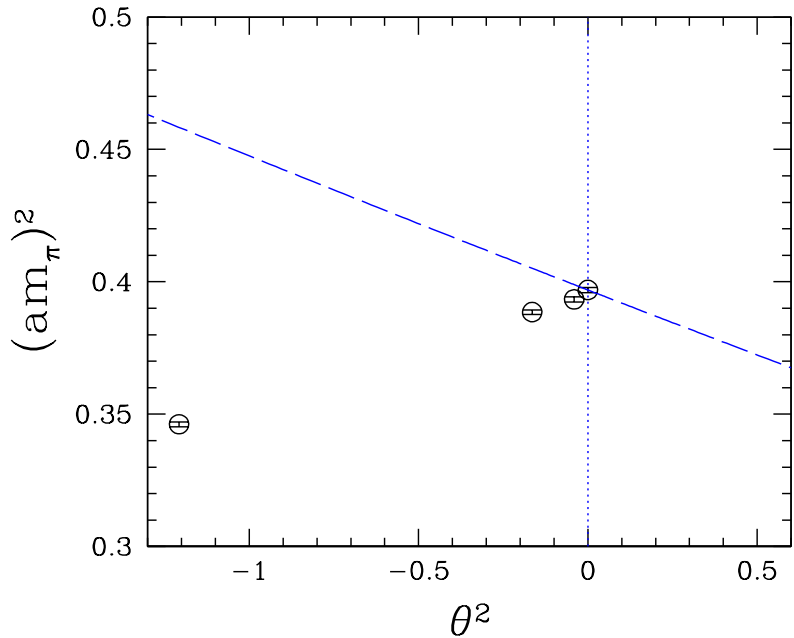,width=11cm,clip=}
\caption{The pion mass squared as a function of $\theta$ squared. The dashed
  line shows the prediction of chiral perturbation theory.}
\end{center}
\end{figure}

A caveat of our calculation is that clover fermions, though $O(a)$ improved,
break chiral symmetry at finite lattice spacing. It is reassuring that $Z_m^S
Z_P \approx 1$, which indicates good chiral properties already. But there is
the potential danger that the remaining $O(a^2)$ corrections will interfere
with the assumed form of the nucleon matrix element (\ref{rw}) and give rise to
systematic errors~\cite{Aoki}. To fully rule that out, we need to repeat the
simulations at smaller lattice spacing.~\footnote{As an independent check for
  lattice artifacts, we have repeated the calculation for zero $\theta$
  angle of the valence quarks on our $\bar{\theta}^I=0.4$ dynamical background
  field configurations and found a nonvanishing result for $d_N^\theta$. The
  result would have been zero in the absence of vacuum insertions of the
  pseudoscalar density~\cite{Guadagnoli}.}  

Finally, we plan to explore a far wider range of $\theta$ values, intrigued by
the results of a recent simulation of the $O(3)$ nonlinear sigma model in two
dimension at imaginary $\theta$~\cite{Alles}. By analytic continuation to real
values of $\theta$ it was possible to detect the phase transition of the model
at $\theta=\pi$ and show that the mass gap vanishes at this point, in
agreement with known results~\cite{Bietenholz}.

\begin{acknowledgments}
We like to thank Wolfgang Bietenholz for carefully reading the manuscript and
%Sinya Aoki and 
Luigi Del Debbio for useful discussions. The simulations of the
background gauge field have been performed on the BlueGene/L at KEK under the
{\it Large Scale Simulation Program 07-14}, while the analysis and computation
of the form factors have been done on the APE computers at DESY Zeuthen. We
thank both institutions for their support. This work is supported in part by
DFG under contract FOR 465 (Forschergruppe Gitter-Hadronen-Ph\"anomenologie)
and 446JAP113/345/0-1 (DFG-JSPS Cooperation Agreement), and by the EU
Integrated Infrastructure Initiative Hadron Physics (I3HP) under contract
RII3-CT-2004-506078. JZ is supported by STFC Grant PP/D000238/1.
\end{acknowledgments}

\end{document}